*Communication*

# Does Double Biofeedback Affect Functional Hemispheric Asymmetry and Activity? A Pilot Study

**Valeriia Demareva [1,*], Elena Mukhina [2], Tatiana Bobro [3] and Ildar Abitov [4]**

[1] Faculty of Social Sciences, Lobachevsky State University of Nizhny Novgorod, 603950 Nizhny Novgorod, Russia;kaleria.naz@gmail.com
[2] Faculty of Social Sciences, Lobachevsky State University of Nizhny Novgorod, 603950 Nizhny Novgorod, Russia; helen_loky@mail.ru
[3] Department of Professional Education, Nizhny Novgorod Institute of Education Development, 603122 Nizhny Novgorod, Russia; btpnn@mail.ru
[4] Institute of Psychology and Education, Kazan State Federal University, 420008 Kazan, Russia; ildar-abitov@yandex.ru
* Correspondence: kaleria.naz@gmail.com; Tel.: +7-904-66-49-13

**Abstract:** In the current pilot study, we attempt to find out how double neurofeedback influences functional hemispheric asymmetry and activity. We examined 30 healthy participants (8 males; 22 females, mean age = 29; SD = 8). To measure functional hemispheric asymmetry and activity, we used computer laterometry in the 'two-source' lead-lag dichotic paradigm. Double biofeedback included 8 min of EEG oscillation recording with five minutes of basic mode. During the basic mode, the current amplitude of the EEG oscillator gets transformed into feedback sounds while the current amplitude of alpha EEG oscillator is used to modulate the intensity of light signals. Double neurofeedback did not directly influence the asymmetry itself but accelerated individual sound perception characteristics during dichotic listening in the preceding effect paradigm. Further research is needed to investigate the effect of double neurofeedback training on functional brain activity and asymmetry, taking into account participants' age, gender, and motivation.

**Keywords:** laterality; brain; hemisphere; asymmetry; activity; biofeedback; EEG

## 1. Introduction

Currently, human beings are overwhelmed with ever increasing information flows so that one speaks about information overload and information stress [1, 2]. In the era of global digitalization and information expansion, it should be studied how modern neurointerfaces may help to optimize humans' physical and psychological state and activate their cognitive resources. Thus, modern cognitive science should develop technologies that could exploit functional brain asymmetry and activity.

### *1.1. Hemispheric Asymmetry*

Currently, the assumption of the dynamic nature of functional hemispheric asymmetry is widely accepted [3].

In general, functional hemispheric asymmetry describes functional difference between symmetric brain structures. Once developed, it is subject to change as a result of compensatory restructuring of structural functional relations caused by lesions. The asymmetry of functional brain relations can be modulated under different manipulations [4]. For example, right-handers may have a greater level of nonspecific activation in the right or in the left hemisphere in a particular state. Some functional states lead to increased elecrophysiological asymmetry up to a statistically significant level. For other functional states, however, such an asymmetry cannot be evidenced. Thus, functional brain asymmetry is determined by a person's functional state, which is defined as the dynamic nature of functional hemispheric asymmetry [4].

In the current study, we will be referring to the dynamic nature of functional hemispheric asymmetry which is expressed in the asymmetry of hemispheric relations conditioned by the distribution of neuronal activity over symmetrical brain structures in the right and in the left hemisphere.

Other terms that are used to refer to what we call functional hemispheric asymmetry are functional connectivity of hemispheres [5], asymmetry of hemispheric network topology [6], functional hemispheric asymmetries [3], hemispheric functional equivalence [7], hemispheric asymmetry of functional brain networks [8].

Selected studies on human participants show that functional differences between the right and the left hemisphere are observed for several cognitive functions (see [9] for an overview). While visual-spatial processing requires right-hemispheric dominance (particularly in fronto-parietal areas [10]), language production and processing (especially for syntax) requires left-hemispheric dominance in adults [11]. Moreover, the degree of hemispheric dominance benefit depends on the handedness [12].

Functional hemispheric asymmetry was broadly studied within a variety of studies on animal models. In general, such studies reveal that the same stimuli are processed by each hemisphere in a different way [9]. For example, functional asymmetry in white matter projections at the junction between midbrain and thalamus is expressed by the fact that the left (contrary to right) hemisphere processes and compares visual objects regardless of their position in the visual field. Asymmetry in the connectivity patterns was found in chicks and pigeons [13]. Furthermore, the results of recent studies in animal models suggest that asymmetries in the connectivity patterns of homologous brain regions are more important for function realization than asymmetries in their size or volume [9]. Artificial removal of asymmetry in symmetrical brain areas is detrimental both for emotional [14] and for cognitive [15] performance.

Dynamic brain asymmetry is highly influenced by 'functional state' (a term widely used within Russian cognitive science and psychophysiology tradition). It refers to a system response of the body which is optimally required by the performed activity [16]. It is shown that the following factors can influence the functional hemispheric asymmetry: mental state [17], emotional arousal [8], depression [18], anxiety [19], mental fatigue [20].

While manual asymmetry characteristics remain more or less constant [21, 22], functional brain asymmetry characteristics are dynamic across the lifespan. For example, it has been shown that frontal asymmetry is not related to handedness [23].

Electrophysiological methods remain the most traditional ones to measure hemispheric asymmetry [24]. One of the ways to quickly and unobtrusively determine functional hemispheric asymmetry is the so-called dichotic listening [25] which generally consists in playing two sound stimuli into two ears. A version of dichotic listening is the so-called computer laterometry [26] which is based on dichotic listening and the precedence effect [27]. They combine into a "two-source" lead-lag paradigm [28].

This technique consists in the presentation of a series of dichotic stimuli with increasing lead-lag delay duration ($\Delta t$). The influence of dichotic stimulation with $\Delta t$ on the magnitude of the afferent response has been clearly demonstrated in a series of electrophysiological studies of anesthetized cats in which the evoked potentials were recorded from symmetrical points of the posterior hills of the quadriplegium during dichotic stimulation with $\Delta t = 0$ and during a 0.1–1.1 ms lead time of one of the ears [29]. Thus, $\Delta t$ is transformed into the hemispheric asymmetry of excitations. Having reached a certain threshold value, hemispheric asymmetry of afferent excitations acts as an unconditional stimulus for the brain, and it triggers reciprocal relationships between paired nerve centers. As a result of reciprocal relations, afferent flows of one half of the brain are completely suppressed (inhibited), and the other half are weakened [30]. So, although laterometry measures subjective sensory space, we can also assess asymmetry of hemispheric relations (distribution of functional activity across hemispheres). The procedure is described in more detail in the Materials and Methods section.

The purpose of this article is to generally explore whether functional hemispheric asymmetry can be changed by external manipulation. We use double biofeedback to induce such a change.

### *1.2. Biofeedback*

To start with, the very specifics of neurointerfaces (biofeedback trainings), through which we can holistically (and unobtrusively) influence the functional state of a person, is briefly considered.

In general, biofeedback is a method of a change of the current human state in a desired direction. Such brain–computer interfaces can capture and process the signals occurring within the human brain, e.g., time-varying EEG signals from individual electrodes [24]. Modern studies demonstrate that EEG is a powerful tool and can be applied in a variety of tasks, e.g., Moore–Penrose pseudoinversion provides reconstruction of the EEG signal which allows to observe differences in brain activity for certain mental tasks. This difference further allows to identify the source that generated a given potential [31].

Biofeedback involves the use of specific hardware–software systems to create real-time visual representations of behavioral or physiological patterns.

Recent research shows that biofeedback is attracting interest not only on the part of physiologists but has also received worldwide approval from specialists in the field of psychology. Such trainings allow to increase the efficiency of human adaptation to various conditions and maximal activation of the inner potential of a human being [32]. An individual can observe such a mapping and modify his or her behavior to achieve a better fit with the reference model. For example, visual–acoustic biofeedback provides visualization of the spectrum of the acoustic speech signal (e.g., [33]), ultrasonic biofeedback provides real-time visualization of tongue shape and movement within the oral cavity (e.g., [33]), to name but a few. Various studies have shown that biofeedback training can facilitate language acquisition in people with speech disorders (e.g., [33—36] as well as in people learning a foreign language [37].

In the current contribution, we suggest that biofeedback can be used as a noninvasive way to change functional hemispheric asymmetry and activity for the following reasons.

(1) Biofeedback can reshape neural networks by increasing their connectivity and neuroplasticity [38, 39]. Globally, biofeedback has been successfully used to modify alpha and gamma brain activity in subjects in order to improve cognitive abilities [40]. Biofeedback has demonstrated its effectiveness in ADHD, autism spectrum disorders, substance use, PTSD, and partial elimination of learning difficulties [41] . Additionally, these procedures can result in long-term changes in the distribution of functional activity across the brain [42], which allows the application of biofeedback in psychiatric rehabilitation [43].

(2) Biofeedback can modify the distribution of functional activity across the cerebral hemispheres.

The technology provides bipolar asymmetry training, which regulates the distribution of frequencies of different intensity. For example, multidirectional changes in the relative EEG intensity in the brain hemispheres can cause cognitive activity and attention processes to be enhanced [44], as well as the improvement of executive functions [45, 46]. Studies have shown that the effectiveness of biofeedback is determined by many factors such as: motivation [47, 48], training [49], respiration [50], and some individual characteristics [51, 52]. In general, by unlocking the body's full potential and mobilizing resources to achieve the goal, one can eliminate excessive psycho-emotional tension and improve the efficiency and rationality of activities not only during biofeedback trainings, but also thereafter.

Most well-studied is biofeedback with the activation of the alpha-rhythm. The findings indicate that the 'alpha state', along with general relaxation, leads to the activation of various types of cognitive activity, in particular—attention [53], memory [54, 55], and even creativity [56]. After biofeedback training, subjects showed improvements in selective attention [57], error detection [55], and a decrease in anxiety [58].

### *1.3. Double Biofeedback*

Double biofeedback is a relatively new technology, which is seen as an effective tool for optimization of the psychophysiological state to execute a wide range of cognitive functions [59]. The fundamental difference of double biofeedback from the one described above is that there is no need for the subject to intentionally optimize their (physiological) reactions on their own by their mapping/visualization. In this case, the loop is closed without conscious participation of the experimental subject—but by their own physiological reactions. The state is optimized faster, and one session is sufficient for a pronounced effect. The optimal measures for feedback lag are established [60]. Double biofeedback is technically a more complicated solution, so it is not used as universally as the one-loop biofeedback.

### *1.4. Biofeedback and Hemispheric Asymmetry*

To our knowledge, there are no studies investigating the dynamics of functional hemispheric asymmetry and activity during biofeedback and double biofeedback. There are some studies investigating the influence of biofeedback on different types of hemispheric asymmetry. The majority of works investigate short-term changes in the asymmetry of alpha-rhythm intensity during biofeedback [61—63] or frontal asymmetry [64]. Individual works study the asymmetry of slow cortical potentials [65]. Since functional hemispheric asymmetry can vary depending on the general level of brain activation, which can be caused by prevention of energy depletion and be compensatory during the functional state change [66], such activation during double biofeedback is seen as an efficient tool to 'manage' functional hemispheric asymmetry and activity.

The purpose of the current study was to reveal the general patterns of double biofeedback influence on functional hemispheric asymmetry and activity.

## 2. Materials and Methods

### *2.1. Participants*

Thirty healthy individuals (8 males, 24 females; mean age = 29; SD = 8) took part in the current study. This group of respondents was chosen because of their accessibility. During the 2020 pandemic, it was difficult to arrange systematic visits of the subjects to the lab to participate in the study. The main goal was to collect a minimum of 30 subjects. The final study sample consisted primarily of social science students (*n* = 21, 70%) who are pursuing a degree in psychology or sociology. Five (17%) were young social science faculty employees. Four (13%) represent students of the medical university and the academy of water transport.

### *2.2. Computer Laterometry*

To investigate functional brain asymmetry, we used computer laterometry which has been developed based on fundamental neurophysiological research and patented procedures of investigating lateral sensor asymmetry [67-70]. Computer laterometry allows to construct various temporal–amplitudinal structures of rectangle sound and noise impulses as well as reaction recording. Stimulus presentation consists in short sound clicks that are given into two ears at registered temporal delays through stereo headphones. In the current research, the virtual acoustic space was constituted by a series of dichotic impulses at a frequency of 3 Hz with the increasing lead-lag delay duration at the rate of 23 microseconds. Sound click intensity was kept constant across participants and did not exceed 40 dB from the monaural hearing threshold with a duration of 50 microseconds. In general, computer laterometry provides the evaluation of lead-lag thresholds for precedence effect components [27] and is not preceded by audiometric examination.

The procedure was as follows. During the training phase, the participants were familiarized with virtual auditory space stimuli. At the beginning of the experimental phase, the participants were requested to give a joystick response when (1) the sound started shifting from the vertex to one of the ears; (2) the sound reached extreme lateralization, i.e., it was clearly heard around one of the ears; (3) there appeared a wholesome image consisting of two independent sounds in two ears (one of them dominant and loud and the other 'echo' sound which is distinct, but quiet). Stimuli were presented first on the left and then on the right-hand side.

To evaluate functional brain asymmetry, basic laterometry parameters were recorded.

1. Δt min L (μs)—lead-lag delay when the virtual auditory space stimulus started shifting from the vertex with left ear advance.
2. Δt min R (μs)—lead-lag delay when the virtual auditory space stimulus started shifting from the vertex with right ear advance.
3. Δt max L (μs)—lead-lag delay at extreme lateralization with left-ear advance.
4. Δt max R (μs)—lead-lag delay at extreme lateralization with right-ear advance.
5. Δt echo L (μs)—lead-lag delay at the 'echo'-effect with left-ear advance.
6. Δt echo L (μs)—lead-lag delay at the 'echo'-effect with right-ear advance.

Functional hemispheric asymmetry coefficients were calculated as follows.

$$K\ min = (\Delta t\ min\ R - \Delta\ t\ min\ L)/(\ \Delta t\ min\ R + \Delta t\ min\ L) \quad (1)$$

$$K\ max = (\Delta t\ max\ R - \Delta t\ max\ L)/(\Delta t\ max\ R + \Delta t\ max\ L) \quad (2)$$

$$K\ echo = (\Delta t\ echo\ L - \Delta t\ echo\ R)/(\Delta t\ echo\ L + \Delta\ t\ echo\ R) \quad (3)$$

$$K\ all = \sqrt{K\ min^2 + K\ max^2 + K\ echo^2} \quad (4)$$

K min denotes the coefficient of hemispheric asymmetry for the first precedence effect. This coefficient reflects the functional dominance of the right (K min > 0) or left (K min < 0) hemisphere in terms of lability. K max denotes the coefficient of hemispheric asymmetry for the second precedence effect. This coefficient reflects the functional dominance of the right (K max > 0) or left (K max < 0) hemisphere in terms of excitability. K echo denotes the coefficient of hemispheric asymmetry for the third precedence effect. This coefficient reflects the functional dominance of the right (K echo > 0) or left (K echo < 0) hemisphere in terms of stability.

It is claimed that there is a 5 to 1 ratio of contralateral to ipsilateral fibers in the auditory system [71]. Thus, with the right ear advance, we evaluate the functional activity of the left hemisphere: with the left ear advance the functional activity of the right hemisphere is estimated [72]. For K min/max/echo > 0, the right hemisphere is dominant on this parameter. For K min/max/echo < 0, then the left hemisphere is functionally dominant on this parameter.

Based on the previous results with regard to tuning functions of multiple neurons at different hierarchical levels of the auditory system and the dynamics of event-related potentials during lead-lag desynchronization processing, approximate individual modules [72] have been defined that reflect the three components of the precedence effect (the start time of the virtual auditory space stimulus shifts from the center, extreme lateralization, and 'echo' [27, 73, 74] as well as characteristics of the virtual auditory space stimuli during dichotic simulation with increasing lead-lag delay duration (Δt min, Δt max, Δt echo, Table 1).

**Table 1.** Reference modules for the precedence effect components.

| Characteristics of Cognitive and Neural Representations | |
| --- | --- |

| Precedence Effect Components | Quantity/ Localization of Virtual Auditory Space Stimuli | The Components of the Auditory Evoked Potential | Auditory System Regions |
|---|---|---|---|
| Δt min | One/ vertex | wave V of the short-latency complex | Stem regions |
| Δt max | One/ maximum lateralization on the advance signal side | N1 of the long-latency complex | Auditory cortex |
| Δt echo | Two/ maximum lateralization on Right/Left side | N1, P2 and late response of the long-latency complex | Frontal, parietal, occipital |

For the first precedence effect component, the lead-lag delay is measured which is sufficient for the perception of the fusion sound to have shifted from the vertex (the center of the inter-ear bend) to the side of advance stimulus presentation (Δt min). That is, lead-lag duration reflects in how far the advance hemisphere is labile, i.e., it can be pre-activated prior to dominating.

For the second precedence effect component, the lead-lag delay that provides maximal lateralization of the virtual auditory space stimulus (Δt max) is measured, that is, the shift towards the extreme side location equivalent to the monaural localization. Thus, Δt max reflects to what extent the advance hemisphere is excitable, i.e., how quickly it can start dominating.

For 'echo', the lead-lag delay is measured which is sufficient to transform two temporally disparate signals into two spatially disparate virtual auditory space stimuli. That is, Δt echo reflects to what extent the advance hemisphere is stable, i.e., how long it can preserve dominance and inhibit the other hemisphere.

Thus, basic laterometry parameters (Δt min, Δt max, Δt echo) are related to hemisphere lability, excitability, and stability.

The lower the Δt min, the higher the lability of the hemisphere that is opposite to the sound shift direction, which reflects the lower activation threshold for neuronal corollaries in the brain stem.

The lower the Δt max, the greater the excitability of the hemisphere that is opposite to the sound shift direction, which reflects the lower activation threshold for neural corollaries in the primary auditory cortex.

The lower the Δt echo, the lower the stability of the hemisphere that is opposite to the sound shift direction, which reflects the shorter time span of neuronal activity in the frontal, parietal, and occipital cortex.

Based on what we have mentioned previously, we can assume, that by comparing Δt min (stands for hemisphere lability), Δt max (stands for hemisphere excitability), and Δt echo (stands for hemisphere stability), with sound replay leading to the right and to the left, we can evaluate functional brain asymmetry in terms of lability, excitability, and stability

Thus, the following two phenomena were looked at with the help of computer laterometry in the current study.

1. Functional hemispheric asymmetry, as reflected in K min, K max, K echo.
2. Functional hemispheric activity, as reflected in Δt min, Δt max, Δt echo and interpreted as lability, excitability, and stability.

### *2.3. Double Biofeedback*

Double biofeedback is a dual loop biofeedback system for monitoring rhythmic brain activity (with vertex electrode) and correcting/optimizing cognitive functions and emotional states. Unlike similar techniques, this technology allows the detection of spectral components of brain biopotentials with high frequency resolution, and in combination with the original resonance stimulation methods, gives an opportunity to detect and analyze individual narrow-frequency EEG oscillators of an individual [59, 60].

This device eliminates the limitations of existing EEG-based biocontrol methods through unique innovations. First, it does not use predetermined, overly broad-band traditional EEG rhythms (theta—4–8 Hz, alpha—8–13 Hz, beta—13–25 Hz, etc.), but automatically detects in real time narrow-frequency EEG oscillators that are characteristic of a particular individual and significant for them. This approach appears quite promising since it takes account of the functional heterogeneity of EEG and the effectiveness of using subjects' narrow EEG frequencies for EEG biofeedback [75]. Secondly, it facilitates teaching a person to self-regulate their state by introducing an additional feedback loop that works automatically simultaneously with the conscious adaptive biocontrol loop. One introduced, f automatic modulation of sensory influences by endogenous rhythms eliminates dependence of the procedure's efficiency on the subject's level of motivation. Thirdly, two feedback channels and an EEG channel that are included into the feedback loop will provide faster influence on the functional activity of hemispheres. The reported advantages of the technology which contribute to its increased efficiency are implemented in a microprocessor device. This device was used for data collection in the current study.

### *2.4. Study Design*

The experiment was administered in several consequent steps.

1. The participant was asked to wear headphones and LED glasses. The headphones were used for laterometry as well as to play double biofeedback sound stimuli.
2. The evaluation of functional hemispheric asymmetry and activity with the help of laterometry.
3. Double biofeedback
   (1) The recording of narrow-frequency components within the pre-defined EEG range (4–20 Hz with 0.1 Hz frequency increase every three seconds) that are dominant for a particular participant. Total recording duration was equal to 480 s.
   (2) General double biofeedback mode where the current amplitude of the respective EEG oscillator from the pre-defined range (4–20 Hz) is transformed into feedback sound signals and the current amplitude of the respective alpha EEG oscillator is used to modulate the intensity of sinusoidal light signals that are generated at the frequency rate of this oscillator. Total recoding duration was equal to 300 sec.
4. The evaluation of the functional hemispheric asymmetry and activity with the help of laterometry.

The study design and procedure were approved by the Ethics Committee of Lobachevsky State University, and all participants provided written informed consent in accordance with the Declaration of Helsinki.

### *2.5. Data Analysis*

To estimate the difference between the parameters of functional brain activity (Δt min, Δt max, Δt echo) before and after double biofeedback, as well as during the sound shift to the left and to the right, T-test for dependent samples was used. To analyze proportional differences, two proportion Z-test was used.

To estimate the differences on the investigated parameters for male and female participants, non-parametric Mann–Whitney test for independent samples was used. To estimate the differences between the parameters of functional brain activity (Δt min, Δt max, Δt echo) within the groups of males and females separately, Wilcoxon matched pairs test was used.

## **3. Results and Discussion**

### *3.1. Functional Hemispheric Asymmetry Dynamics*

The analysis of functional hemispheric activity was performed by three criteria: 1) hemispheric dominance (by lability, excitability, stability) before and after the biofeedback; 2) statistical significance of functional brain asymmetry coefficient changes by lability, excitability, and stability; 3) individual analysis of the dynamics of hemispheric activity dominance as to lability, excitability, stability.

The distribution of hemispheric dominance by lability, excitability, and stability before and after the biofeedback is shown in Figure 1a–c.

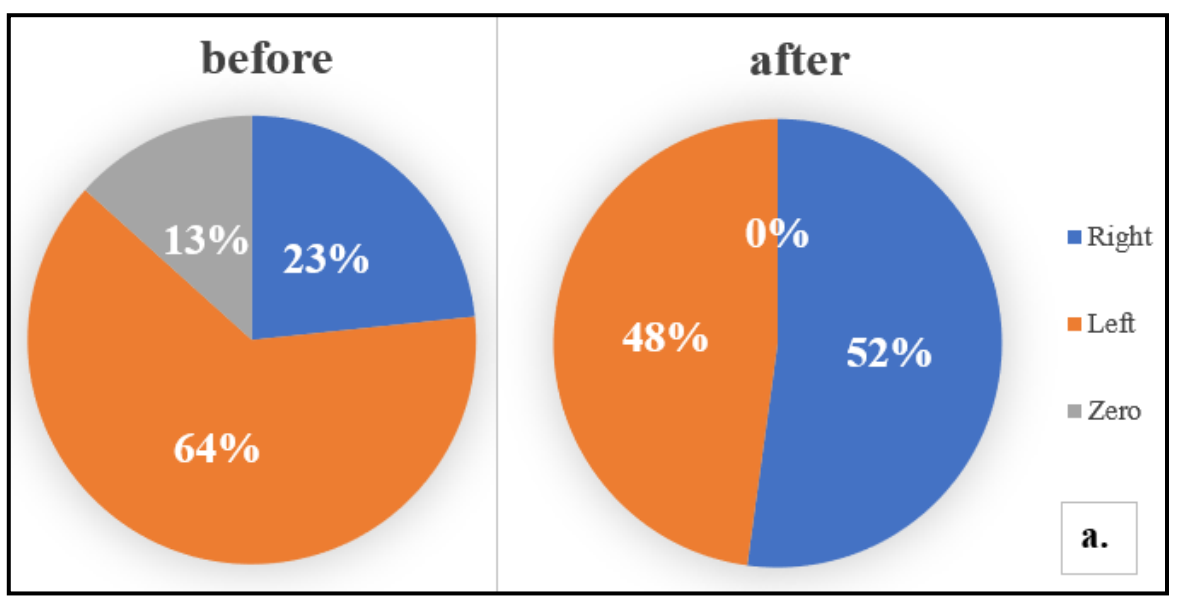

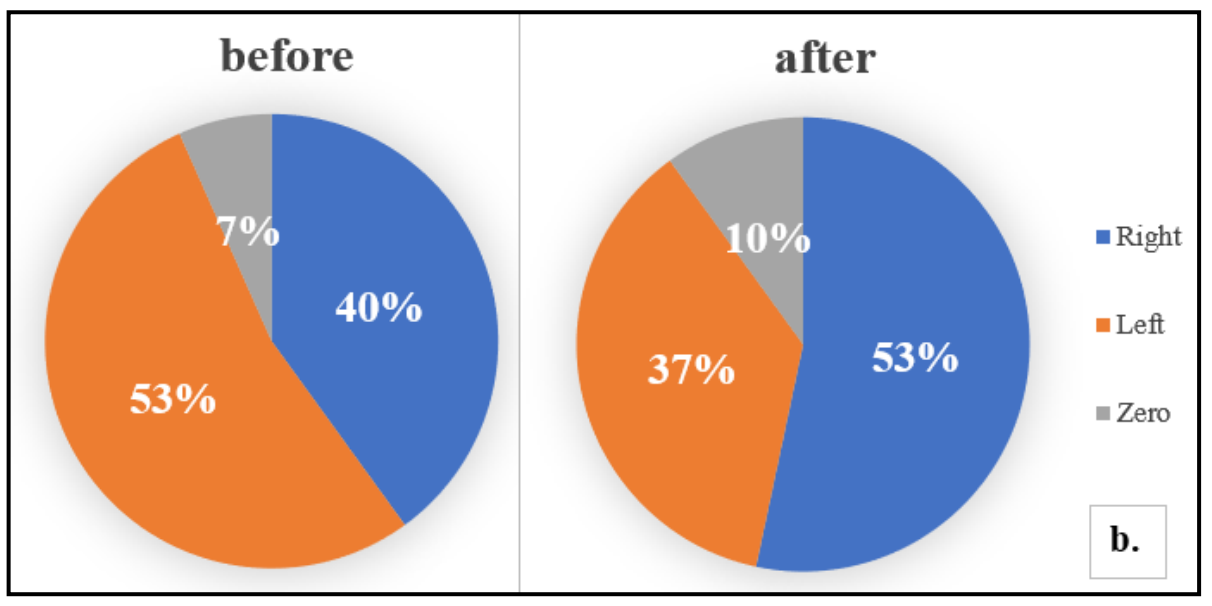

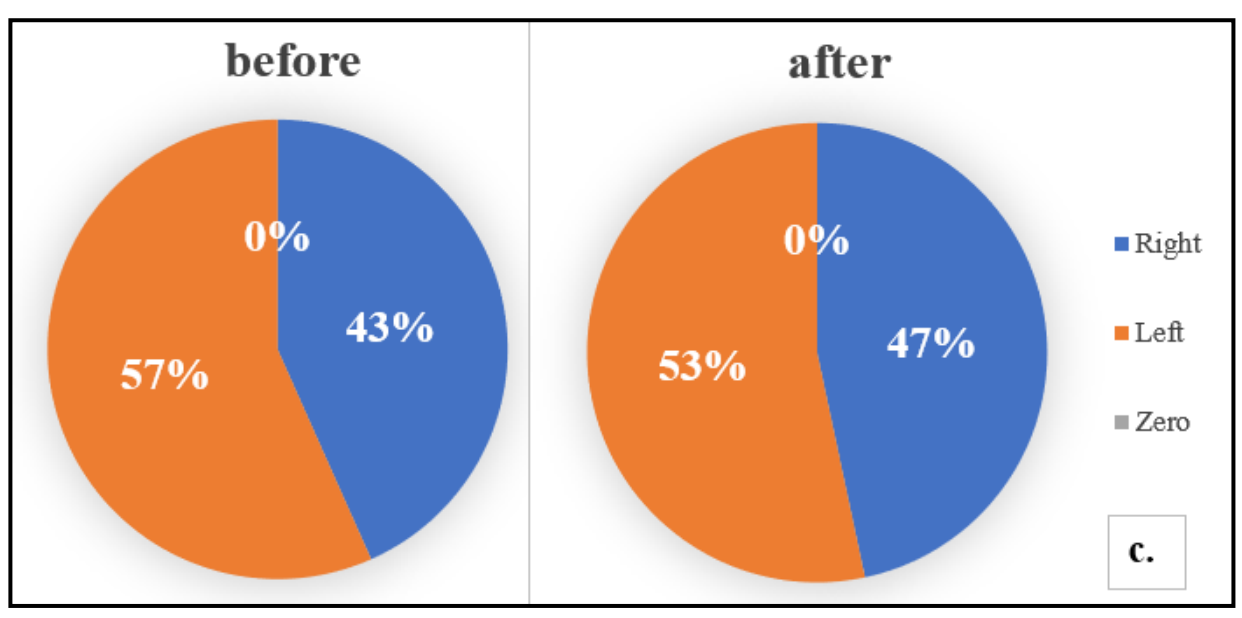

**Figure 1.** Dominant hemisphere by lability (**a**), excitability (**b**), and stability (**c**) before and after biofeedback ('Right'—K min/max/echo > 0, 'Left'—K min/max/echo < 0, 'Zero'—K min/max/echo = 0).

The results of statistical analysis of functional brain asymmetry coefficients before and after feedback are presented in Table 2.

The sample is relatively evenly distributed by functional activity dominance in terms of excitability and stability both before and after double biofeedback. As to lability, the majority of participants were left hemisphere dominant prior to double feedback (19 participants—64%). Their prevalence is statistically significant (two proportion Z-test ($p <$ 0.01)). Such an effect is still to be explained. It can be connected with predominantly evening testing sessions, yet the influence of the time of testing has not been previously investigated and remains to be investigated in the future.

**Table 2.** Functional hemispheric asymmetry coefficients before and after double biofeedback training (means and standard deviations), t-test, and *p*- values.

| | Mean | | *SD* | | *t* | *p* |
|---|---|---|---|---|---|---|
| | **Before** | **After** | **Before** | **After** | | |
| K min (lability) | −0.064 | −0.018 | 0.136 | 0.136 | −1.3 | 0.197 |
| K max (excitability) | −0.011 | 0.011 | 0.098 | 0.099 | −0.9 | 0.372 |
| K echo (stability) | −0.020 | 0.005 | 0.125 | 0.096 | −0.9 | 0.373 |
| K all | 0.203 | 0.179 | 0.078 | 0.070 | 1.1 | 0.281 |

Functional hemispheric asymmetry coefficients did not change before and after the double biofeedback training. So, globally functional brain asymmetry did not change as to any of the components (lability, excitability, or stability). It can be rooted in more complex mechanisms underpinning brain activity during double biofeedback training. Secondly, it can be caused by other factors that have not been accounted for in the current study.

We then conducted individual analyses of functional hemispheric asymmetry change for the whole sample. The data are summarized in Table 3.

**Table 3.** Changes in the dominance of hemispheric activity by lability, excitability, stability*.

| **Δ t min (Lability)** | | **Δ t max (Excitability)** | | **Δ t echo (Stability)** | |
|---|---|---|---|---|---|
| Left -> Right | Right -> Left | Left -> Right | Right -> Left | Left -> Right | Right -> Left |
| 7 | 3 | 8 | 5 | 7 | 6 |
| Zero -> Right | Zero -> Left | Zero -> Right | Zero -> Left | Zero -> Right | Zero -> Left |
| 2 | 2 | 1 | 1 | 0 | 0 |
| Change to Right | Change to Left | Change to Right | Change to Left | Change to Right | Change to Left |
| 9 | 5 | 9 | 6 | 7 | 6 |
| No change | | No change | | No change | |
| 16 | | 15 | | 17 | |

* 'Left'—the laterometry parameter indicates greater activity of the left hemisphere (Δ t min/Δ t max/Δ t echo). 'Right'—the laterometry parameter indicates greater activity of the right hemisphere (Δ t min/Δ t max/Δ t echo). 'Zero'—no asymmetry on the measured parameter (Δ t min/Δ t max/Δ t echo)).

Half of the participant sample did not experience any functional hemispheric activity change on any of the components—lability, excitability, stability (53%, 50%, 57%, respectively). Among those who experienced asymmetry change, there is a tendency to the rightward shift. This tendency is more pronounced for lability. Among 14 people who experienced asymmetry inversion, in 9 (64%), it changed to the right, yet this change is not statistically significant (two proportion Z-test $p = 0.14$). Thus, a greater sample is needed to attest this tendency.

In general, functional hemispheric asymmetry change in the current experiment supports previous claims with regard to its dynamic nature and its possible modification under the influence of different factors.

### *3.2. Functional Hemispheric Activity Dynamics*

The analysis of functional hemispheric activity was performed with regard to (1) the direction of the sound shift (left vs. right); (2) time point (before vs. after double biofeedback).

No difference in Δ t max and Δ t echo was observed for sound shift to the left or to the right at any stage of the experiment. Only one statistically significant difference was observed in Δ t min before double biofeedback—see Figure 2.

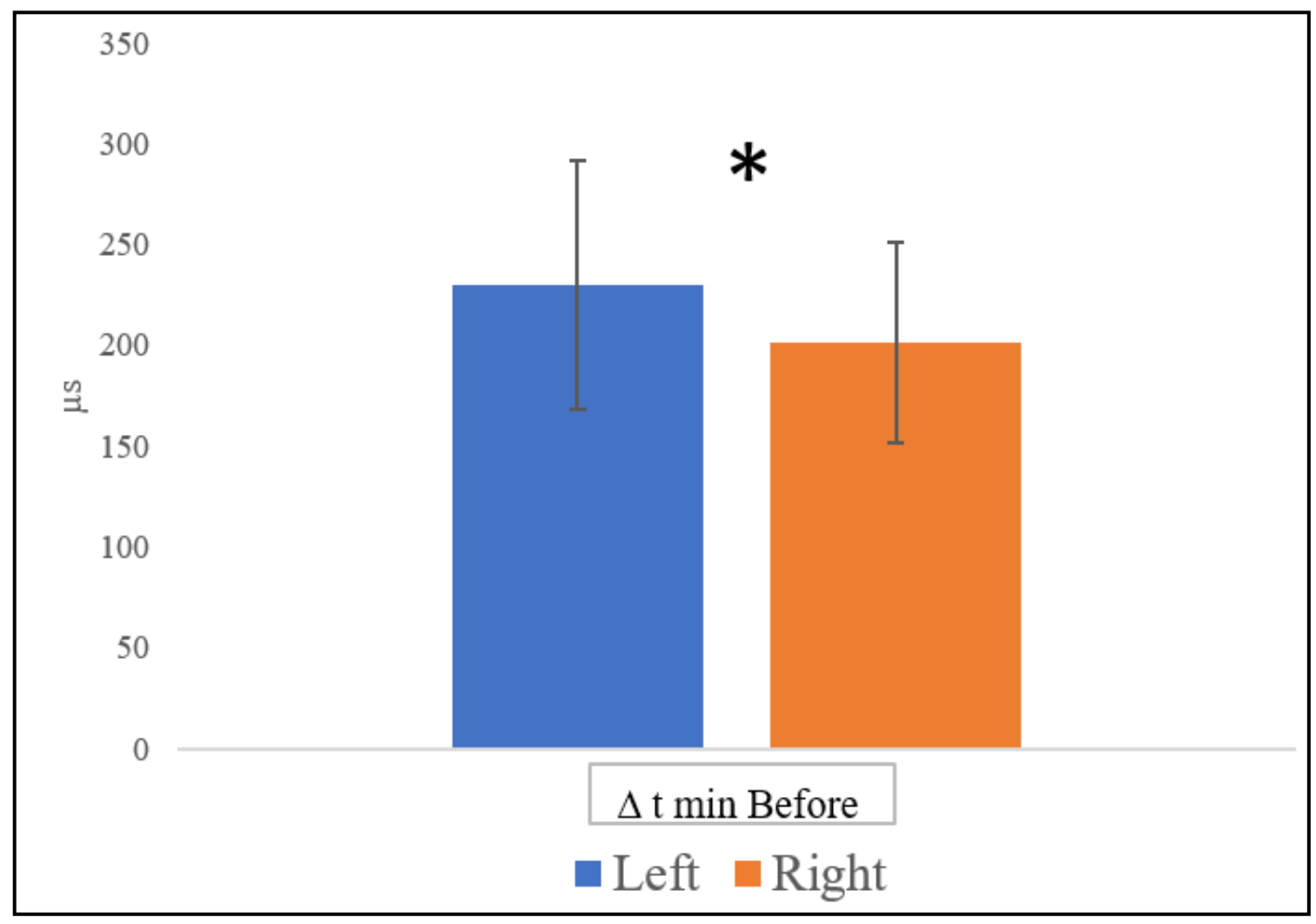

**Figure 2.** Mean values for Δ t min left and right before double biofeedback (vertical bars denote standard deviations; *—$p < 0.05$—T-test for dependent samples).

Thus, prior to double feedback, there was a statistically significant difference in hemispheric lability: Δ t min before left is greater than Δ t min before right ($t = 2.6$, $p < 0.05$). It is in line with the general distribution of the sample on initial functional hemispheric asymmetry—see Figure 1a.

Δ t min left, Δ t echo left, and Δ t echo right were significantly reduced after double feedback training—see Figure 3.

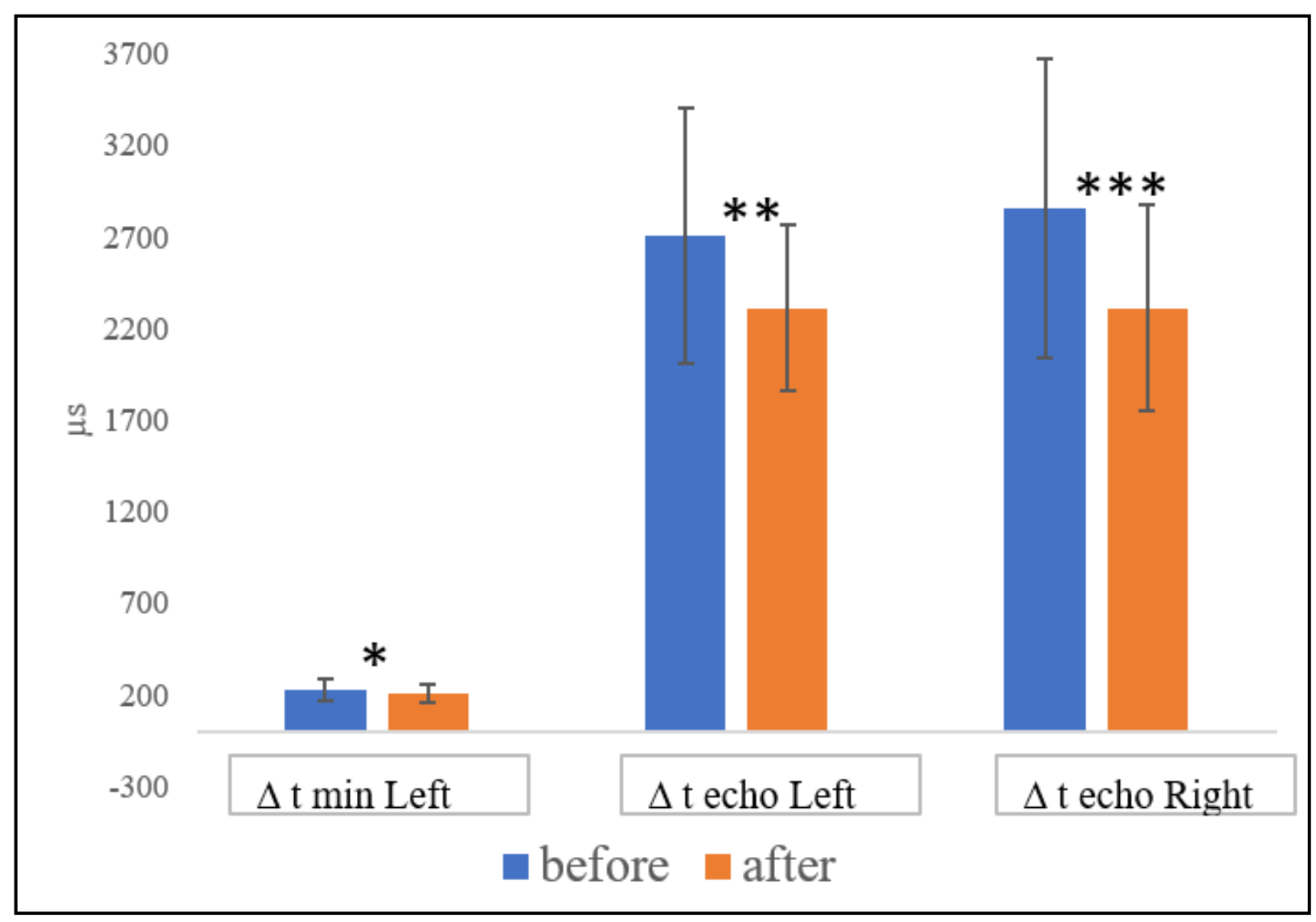

**Figure 3.** Mean values for Δ t min left, Δ t echo left, and Δ t echo right before and after double biofeedback (vertical bars denote standard deviations; *—$p < 0.05$, **—$p < 0.01$, ***—$p < 0.001$—T-test for dependent samples).

Thus, we found that there was a statistically significant acceleration of information processing in the brain stem of the right hemisphere as well as in the frontal, parietal, and occipital lobes of the right and the left hemispheres.

Stability reduction reveals that one of the hemispheres is quick to give preferential processing of sound information over to the other hemisphere. Possibly, double biofeedback contributes to deactivation of reciprocal inhibition. One can say that double biofeedback contributes to general brain activation across quite a number of dimensions, which is in line with what has been shown previously [53—58].

Such an effect may seem to be related to timing or learning. Yet, functional brain asymmetry evaluation with the help of computer laterometry was also used in other projects in a variety of contexts. The tendency to lability and stability reduction has not been reported (e.g., [76]). Future research should take it into account by introducing a control group of participants whose functional hemispheric asymmetry is screened at the beginning and at the end of the double biofeedback time window. Another non-biofeedback treated control group, who is just listening to music and experiencing LED flashes which are not modulated by EEG, is also needed.

### *3.3. Gender Effects in Functional Hemispheric Asymmetry and Activity*

Mann–Whitney test did not show any significant differences between male and female participants on any of the parameters. The most 'significant' difference was found for Δ t min (lability) left before biofeedback (U = 52, $p$ = 0.096). Males seem to display smaller values for right hemisphere lability than women before biofeedback.

Further analysis showed that for males, Δt min before right is smaller than Δt min before left stimulus presentation—see Figure 4.

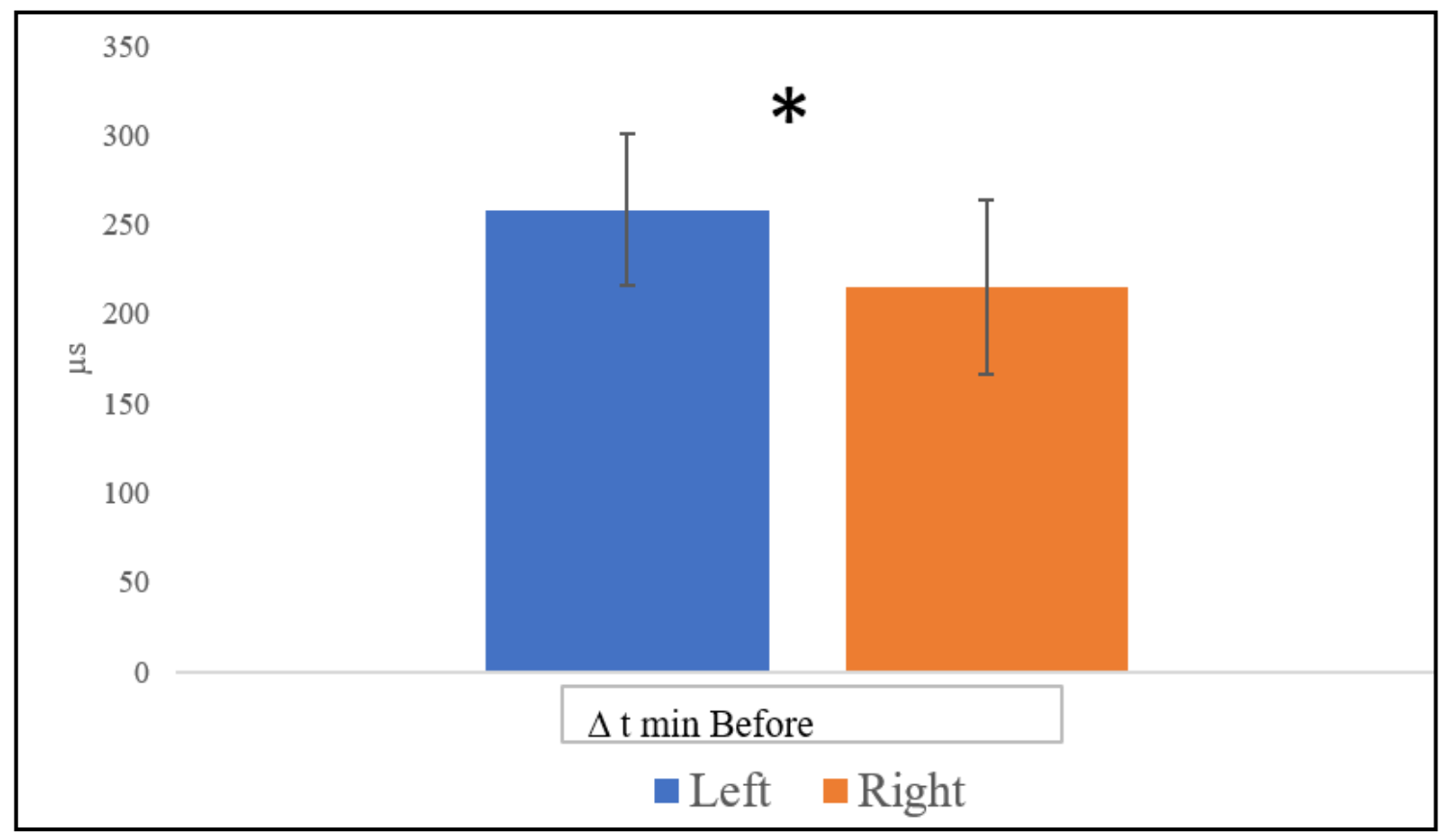

**Figure 4.** Mean values for Δ t min Left before double biofeedback (vertical bars denote standard deviations; *—$p < 0.05$—Wilcoxon matched pairs test).

Thus, in general, men seem to have better values for the left than for the right hemisphere lability. This can be interpreted in a way that they have a tendency for left hemisphere functional dominance for lability—their left hemisphere can be quicker pre-activated prior to dominating than the right one. To study this aspect further, a gender-balanced sample is needed. The goal of the current pilot investigation was to trace basic tendencies in hemispheric asymmetry and activity dynamics before and after double biofeedback training. Gender difference in functional brain asymmetry in its various manifestations is being widely discussed in the literature (see [77] for review).

## 4. Conclusions

1. Double biofeedback did not alter functional brain asymmetry on any of its characteristics—lability, excitability, or stability.
2. An increase in the right hemisphere lability and a decrease in the stability of both hemispheres were evidenced after double biofeedback training.
3. Further research on double biofeedback is needed with different control groups, varying experimental conditions in treatment groups, as well as controlling for gender, age, and time of testing.
4. In the future, we plan to apply the results of the study on the effect of double biofeedback to determine optimal brain states for foreign language acquisition. Since functional hemispheric activity is related to the success of Russian-speaking students in mastering English [76], and since the pilot study revealed variability in this activity before and after double biofeedback, going forward, we will be able to select such double biofeedback protocols, which will be optimal for language acquisition.

5. Computer laterometry itself is a useful and convenient tool for studying functional hemispheric asymmetry and activity. It seems important to conduct a separate study on brain activity during computer laterometry. It can be achieved with modern quantitative EEG.

**Author Contributions:** V.D.—experiment design and methodology, experiment administration, data analysis and preparation of first draft; E.M.—assistance with experiment administration and data collection; T.B. and I.A.—literature review, editing and proofreading the final draft. All authors have read and agreed to the published version of the manuscript.

**Funding:** This research was funded by Lobachevsky State University, grant number H-456-99_20-2 ('Evaluation of the effectiveness of neurobiofeedback to optimize the functional state in learning a foreign language').

**Institutional Review Board Statement:** The study was conducted according to the guidelines of the Declaration of Helsinki, and approved by the Institutional Review Board of Faculty of Social Sciences of Lobachevsky State University (protocol №2 from 08.09.2020).

**Informed Consent Statement:** Informed consent was obtained from all subjects involved in the study.

**Data Availability Statement:** The data presented in this study are available on request from the corresponding author. The data are not publicly available due to their containing information that could compromise the privacy of research participants.

**Acknowledgments:** The authors would like to thank Alexander Fedotchev and Alexander Bondar for providing the double biofeedback device and Sofia Polevaya for explaining technological specificities of the device in context of this study.

**Conflicts of Interest:** The authors declare no conflict of interest.